\documentclass[usenatbib,usegraphicx]{mn2e}

\title[Spectrophotometry of $\epsilon$ Cha cluster members]
{Spectrophotometric properties of pre-main sequence stars: the $\epsilon$ Chamaeleontis cluster}

\author[A-R. Lyo et al.]
{A-Ran Lyo,$^{1}$\thanks{E-mail: arl@kasi.re.kr (ARL); w.lawson@adfa.edu.au (WAL);
bessell@mso.anu.edu.au (MSB)}
Warrick A. Lawson$^{2}$ and M. S. Bessell$^{3}$\\
$^{1}$Korea Astronomy and Space Science Institute, 61-1, Hwaam-dong, Yuseong-gu, Daejeon 305-348, Korea\\
$^{2}$School of Physical, Environmental and Mathematical Sciences, University of
New South Wales, Australian Defence Force Academy,\\ Canberra, ACT 2600, Australia\\
$^{3}$Research School of Astronomy and Astrophysics, Institute of Advanced Studies,
The Australian National University, Cotter Road,\\ Weston Creek ACT 2611, Australia}

\begin{document}

\date{Accepted ........................ Received ........................}

\pagerange{\pageref{firstpage}--\pageref{lastpage}}
\pubyear{2008}

\maketitle

\label{firstpage}

\begin{abstract}

We present a study of flux-calibrated low-resolution optical
spectroscopy of ten stars belonging to eight systems in the $\sim
5$ Myr-old $\epsilon$ Chamaeleontis ($\epsilon$ Cha)
pre-main-sequence (PMS) star cluster. Using synthetic broadband
colours, narrow-band continuum, atomic and molecular lines derived
from the spectra, we compare the $\epsilon$ Cha stars to a
slightly older PMS cluster, the $\approx 8$ Myr-old $\eta$ Cha
cluster, and to main-sequence dwarfs. Using synthetic {\it VRI\,}
colours and other indices, we find that the relationship between
broadband colours and spectroscopic temperature indicators for
$\epsilon$ Cha cluster members is indistinguishable from that of
Gyr-old dwarfs. This result is identical to that found earlier in
$\eta$ Cha. Gravity-sensitive line indices place the cluster
between the giant and dwarf sequences, and there is clear evidence
that $\epsilon$ Cha stars have lower surface gravity than $\eta$
Cha stars. This result is consistent with $\epsilon$ Cha being the
slightly younger PMS association, a few Myr younger according to
the Hertzsprung-Russell (HR) diagram placement of these two
clusters and comparison with PMS evolutionary grids. Late M-type
$\epsilon$ Cha cluster members show a $B$-band flux excess of
$\approx 0.2$ mag compared to observations of standard dwarfs,
which might be related to enhanced magnetic activity. A similar
level of excess $B$-band emission appears to be a ubiquitous
feature of low mass members of young stellar populations with ages
less than a few hundred Myr, a very similar timescale to the PMS
phase of elevated relative X-ray luminosity.

\end{abstract}


\section{Introduction}

Analysis of flux-calibrated low-resolution optical spectra is a
useful method for investigating the physical properties of stars
in comparison with spectra of standard stars with well-defined
properties. For nearby stars and stellar associations, such
observations are readily obtained even with telescopes of modest
aperture (e.g., Bessell 1991; Lyo, Lawson \& Bessell 2004). Using
synthetic broadband colours, narrow-band continuum, atomic and
molecular line indices derived from calibrated spectra, we can
study temperature, surface gravity and metallicity effects, e.g.
Lyo et al. (2004) showed that the relationship between colours and
spectroscopic temperature indicators for the $\approx 8$ Myr-old
$\eta$ Cha cluster was indistinguishable from that for Gyr-old
disk dwarfs, although the spectra showed some evidence for higher
metallicity and clear evidence for lower gravity. The latter is a
consequence of the elevated location of these stars, several mag
above the zero main sequence in the HR diagram. The $\eta$ Cha
stars also displayed, like other young stellar populations, a
$B$-band flux excess attaining $\approx$ 0.2 mag for late-M
cluster members.

By studying PMS clusters of different isochronal ages, we can use
spectrophotometric techniques to address various stellar
evolutionary issues, e.g. is there evidence in the spectroscopy
for differences in the temperature-spectral type sequence as a
function of age? Do gravity-dependent features scale with age as
stellar low mass stars descend their Hayashi tracks? Is the
observed $B$-band excess in late-M PMS stars a general property of
young stellar populations, and does it vary as a function of
spectral type and age?

In this paper, our target group is the young $\epsilon$ Cha
cluster associated with the early-type system $\epsilon$ Cha AB
and the codistant, comoving star HD104237 (DX Cha). HD104237 is
the nearest-known Herbig Ae star and forms at least a quintet with
low mass companions based upon {\it Chandra X-ray Observatory}
observations, optical/infrared imaging and spectroscopic study,
some of which are likely multiple themselves (Feigelson et al.
2003; Grady et al. 2004). HD104237A itself is a spectroscopic
binary with a K3 companion in an eccentric 20-d orbit (B\"ohm et
al. 2004). The group is nearby ($d \approx 114$ pc), compact in
extent ($\approx 1$ pc) and sparsely populated, containing eight
stellar systems in its central region with stars ranging in
spectral type from B9 to M5. HR diagram placement of the members
ages the group at $3-5$ Myr (Feigelson et al. 2003) when compared
to Siess, Dufour \& Forestini (2000) evolutionary tracks, and
$\sim 6$ Myr (Luhman 2004) when using a mix of Palla \& Stahler
(1999), Baraffe et al. (1998) and Chabrier et al. (2000) tracks.
Both the HR diagram comparisons of Feigelson et al. (2003) and
Luhman (2004) assign younger ages to the early-type members of
$2-3$ Myr. This discrepancy in age with spectral type might be a
consequence of nearby low-mass companions elevating the
luminosities of the early-type stars, or might serve to highlight
differences between model isochrones and observational isochrones,
or indicate a genuine age difference between the high-mass and
low-mass stars in this cluster. Kinematic study indicates the
group is another out-lying population of the Oph-Sco-Cen OB
association (Feigelson et al. 2003), following groups and
associations such as $\eta$ Cha, the TW Hya association, and the
$\beta$ Pic moving group.

The $\epsilon$ Cha cluster presents itself then, as a slightly
younger analogue of the $\eta$ Cha star cluster (Mamajek, Lawson
\& Feigelson 1999, 2000), an $\approx 8$ Myr-old compact, sparse
PMS group that was studied using spectrophotometric techniques by
Lyo et al. (2004). The ages of PMS clusters derived by HR diagram
comparisons using competing evolutionary grids are unreliable;
differences can exceed several Myr at $\sim 10$ Myr (see fig. 8 of
Lyo et al. 2004). However, comparison between PMS groups using
techniques such as spectrophotometry ought to reveal differences
that can be interpreted as age-related trends. With an age
difference of only a few Myr, comparison between the $\epsilon$
Cha and $\eta$ Cha groups is a demonstration of the sensitivity of
such methods to rank in age various PMS populations.

\section{Observations and data reduction}

\subsection{Calibration of the spectra}

We obtained low-resolution spectra of ten members of the
$\epsilon$ Cha star cluster during 2005 March using the 2.3-m
telescope and double beam spectrograph (DBS) at Siding Spring
Observatory (SSO). Cluster members range in spectral type from B9
to M5 (Feigelson et al. 2003; Luhman 2004). For comparison with
the early-type cluster members, we obtained DBS spectra of several
early-type B- and A-type stars from the Bright Star Catalogue. For
comparison with the late-type members, we obtained DBS spectra of
main-sequence F- to M-type dwarfs from the Gliese catalogue (Table
1). Bessell (1990) lists Gliese stars with photoelectric
photometry and spectral types.

In the blue arm of the DBS, the 300B (300 line\,mm$^{-1}$) grating
gave a 2-pixel resolution of 4 \AA\, with coverage from
$\lambda\lambda 3500-6000$ \AA.  In the red arm, the 158R (158
line\,mm$^{-1}$) grating gave a 2-pixel resolution of 8 \AA\, from
$\lambda\lambda 5000-11000$ \AA.  The blue and red spectra were
obtained simultaneously in a pseudo-spectrophotometric mode, with
the slit width set to maximize the spectral resolution and
oriented to the parallactic angle to eliminate differential
refraction.

The spectra were first reduced using dome flats, bias frames and
Cu-Ar arc frames after removing cosmic rays (using the automatic
{\tt IRAF} {\tt cosmic ray} routine, or manually using {\tt
imedit}) and making use of standard {\tt IRAF} library routines
such as {\tt ccdproc} and {\tt apall}. Further details of the
spectroscopic reduction process, in particular the removal of
telluric absorption features, are described in Lyo et al. (2004).
The reduced DBS flux-calibrated spectra of members of the
$\epsilon$ Cha cluster are shown in Fig. 1\footnote{A file
containing these spectra can be downloaded from
http://www/mso.anu.edu.au/$\sim$bessell/FTP/EpsCha\_spectra/.}.

\begin{figure*}
\begin{center}
\includegraphics[width=185mm]{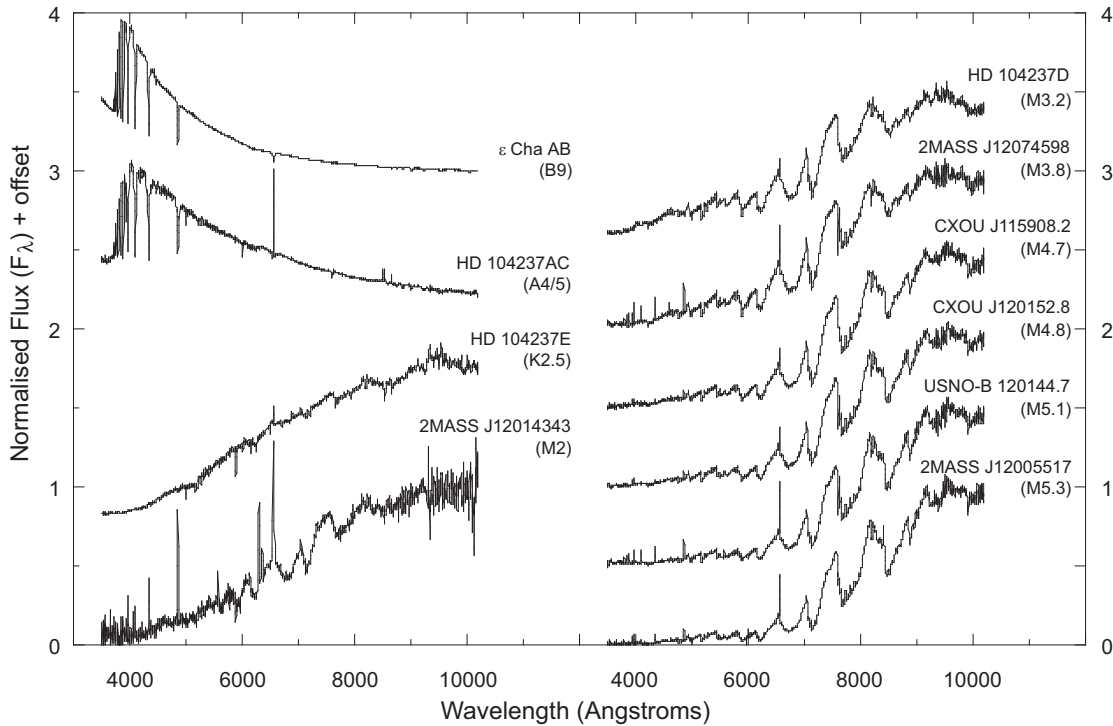}
\caption{SSO 2.3-m/DBS flux-calibrated spectra of members of the
$\epsilon$ Cha cluster. $\epsilon$ Cha AB is a late B-star with an
A-type companion with separations reported between 0.45$''$ and
1.9$''$. HD 104237A is a Herbig Ae star with several low-mass
companions at separations between 1$''$ and  15$''$. The nearest
companions of these systems are not resolved individually in the
DBS observations due to their close separation and the brightness
of their primaries. The strong H$\alpha$ emission line of 2MASS
J12014343-7835472, with an equivalent width of 80 \AA\,, is shown
truncated in this figure so as not to overlap with the spectrum of
HD104237E.} \label{Fig1}
\end{center}
\end{figure*}

\subsection{Derivation of the synthetic colours}

We then obtained synthetic colours from the flux-calibrated
spectra of the $\epsilon$ Cha cluster stars and the Gliese dwarfs.
First, we obtained instrumental colours by integrating the spectra
with normalized filter response functions in the Cousins {\it
BVRI\,} system.  Using published photoelectric photometry for the
Gliese stars from Bessell (1990), reproduced in Table 1, we then
derived transformation equations to convert the instrumental
colours to the Cousins system. For all of the colours, the
transformation equation was a linear function with a regression
coefficient $R \approx 1$. Comparison of the synthetic colours
derived from the spectra, to the colours obtained from the
photoelectric photometry, shows the scatter of the colours
averages 0.05 mag (1$\sigma$).

We adopted the $(R-I)$ colour as the primary broadband colour for
characterizing the $\epsilon$ Cha stars since a single epoch
$(R-I)$ measurement appears to better describe the stars than
other colours more-affected by variability (Bouvier et al. 1993;
Lawson et al. 2001, 2002), and also because the $(R-I)$ colour has
been employed before to characterize the late-type dwarf spectral
sequence (Bessell 1991).

\begin{table*}
\centering \caption{$V$ mag, colours and spectral types of F- to
M-type dwarfs from the Gliese catalogue (Bessell 1990). The
spectral types listed in the last column are derived using the
spectral type-colour conversions of Kenyon \& Hartmann (1995) for
$F - K$ spectral types, and Bessell (1991) for $M$ spectral
types.} \label{Table1}
\begin{tabular}{@{}lrrrllllc@{}}
\hline Gliese & $\alpha_{1950}$ & $\delta_{1950}$ & $V$ &
$(B-V)$
     & $(V-R)$ & $(R-I)$ & $(V-I)$ & Sp. Type \\ \hline
GL 385   & 10 11 08&--84 50.1&10.28&0.337&0.226&0.210&0.431&F1.2\\
GL 297   & 08 05 21&--04 22.6&11.22&0.380&0.216&0.238&0.452&F1.2\\
GL 404   & 10 51 12&--44 08.7&8.08&0.520&0.300&0.320&0.620&F7.2\\
GL 539.1 & 14 03 48&--74 36.9&6.03&0.580&0.330&0.320&0.650&F8.6\\
GL 550   & 14 24 10&--51 42.6&7.84&0.700&0.365&0.345&0.710&G2.7\\
GL 500   & 13 07 00&--21 55.3&7.34&0.730&0.385&0.370&0.755&G5.3\\
GL 511   & 13 23 56&--24 02.0&8.78&0.930&0.523&0.465&0.988&K1.9\\
GL 545.1 & 14 18 18&--40 09.9&9.00&1.108&0.641&0.513&1.152&K4.1\\
GL 340.3 & 09 18 16&--05 32.4&9.07&1.156&0.680&0.568&1.249&K4.4\\
GL 489   & 12 55 07&--14 11.6&9.13&1.120&0.674&0.585&1.256&K4.4\\
GL 542.2 & 14 16 20&--06 22.1&9.09&1.316&0.799&0.711&1.508&K6.3\\
GL 296   & 07 58 15&--39 53.5&9.65&1.350&0.823&0.765&1.589&K6.8\\
GL 334   & 09 04 20&--08 36.5&9.50&1.470&0.914&0.905&1.820&M0.2\\
GL 508.3 & 13 21 07&--13 46.8&11.79&1.410&0.900&0.987&1.885&M0.5\\
GL 341   & 09 20 42&--60 04.5&9.49&1.502&0.941&1.008&1.949&M0.9\\
GL 433   & 11 32 58&--32 15.1&9.84&1.521&1.000&1.150&2.154&M2.0 \\
GL 333   & 08 57 54&--47 15.0&12.18&1.560&1.015&1.190&2.208&M2.2\\
GL 298   & 08 08 42&--52 49.7&11.81&1.515&1.083&1.384&2.473&M3.0\\
GL 543   & 14 16 36&--07 03.8&13.48&1.622&1.095&1.392&2.486&M3.0\\
GL 300   & 08 10 31&--21 23.5&12.16&1.640&1.257&1.648&2.902&M4.1\\
GL 514.1 & 13 27 29&--08 26.6&14.33&1.680&1.320&1.720&3.040&M4.3\\
GL 473   & 12 30 51&  09 17.6&12.49&1.840&1.555&1.960&3.520&M5.2\\
GL 406   & 10 54 06&  07 19.2&13.53&1.990&1.856&2.177&4.032&M5.9\\
 \hline
\end{tabular}
\end{table*}

\section{Temperature sequence with age}

Greene \& Lada (1997) suggested that the effective temperature
scale of PMS stars is likely to be similar to that appropriate for
dwarfs or subgiants since the surface gravities of PMS stars are
more like those of main-sequence dwarfs, than those of giants.
However, Luhman (1997) suggested that the correct T Tauri
temperature scale should be constrained between that of dwarfs and
giants since the surface gravity of T Tauri stars appears to be
intermediate between that of dwarfs and giants.

In Lyo et al. (2004), we concluded that the relationship between
colours and spectroscopic temperature indicators for members of
the $\approx 8$ Myr-old $\eta$ Cha star cluster is
indistinguishable from that of the Gyr-old main-sequence dwarfs.
Accordingly, in that paper we adopted dwarf colour-spectral type
conversions for the $\eta$ Cha stars. This result was obtained
through studies of the broad-band colours, narrow-band continuum
and temperature-sensitive spectral indices, using spectra obtained
with the same telescope and instrument as we have used to obtain
observations of members of the $\epsilon$ Cha star cluster.

In the following sub-sections, we investigate how spectroscopic
temperature indicators behave in a younger, $\sim 5$ Myr-old, PMS
star cluster, in comparison to the results obtained previously for
$\eta$ Cha.

\subsection{Broad-band synthetic colours}

In Fig. 2, we show colour-colour diagrams for $\epsilon$ Cha
cluster stars and Gliese dwarfs produced using synthetic colours
derived from our calibrated spectra. In these panels, the solid
lines represent the locus of B0 -- M6 dwarfs in each of these
colour-colour planes, using photoelectric photometry compiled by
Kenyon \& Hartmann (1995) and Bessell (1991). We generally see
very close agreement between the synthetic colours of both groups
of stars and the locus of main-sequence colours. In particular,
the Gliese dwarfs we observed have colours that are well-matched
to the dwarf photometric sequences. This is important for our
later discussion, as we will compare indices derived from the
spectra of the Gliese stars, to those derived from the spectra of
the $\epsilon$ Cha stars.

In Fig. 2(a) the principal difference of note is the lower $(B-V)$
colour seen in low mass $\epsilon$ Cha stars compared to the dwarf
sequence. This is indicative of a $B$-band excess of $\sim 0.2$
mag in the PMS stars, since their $(V-R)$ colours are
indistinguishable from the dwarf sequence. We further discuss the
$B$-band excess in Section 5.

Three other $\epsilon$ Cha stars show deviations from the dwarf
sequence in one or more of the panels. Filled diamonds represent
the observed colours of HD104237E, whereas open diamonds show
extinction-corrected colours assuming $A_V = 1.8$ mag (Feigelson
et al. 2003). The extinction-corrected colours show good agreement
with the dwarf sequence. Filled squares represent 2MASS
J12014343-7835472; the abnormal colours (and spectrum; see Fig. 1)
of the star might be due to the star being observed through an
edge-on disk (Luhman 2004). We confirm the youth of the star
through an observed high level of H$\alpha$ emission
($EW$(H$\alpha$) $\sim$ 80 \AA) and 6300, 6360 \AA\, [OI]
forbidden emission (Fig. 1). Filled triangles represent HD104237D;
the apparent blue excess in the observed colours of the star is
due to contamination from its Herbig Ae primary HD104237A, 8 mag
brighter in $V$-band than HD104237D and only 10$''$ distant. Other
late-type $\epsilon$ Cha stars show no evidence for reddening, as
they display broadband colours consistent with their spectral type
(Feigelson et al. 2003; Luhman 2004).

\begin{figure}
\begin{center}
\includegraphics[width=70mm]{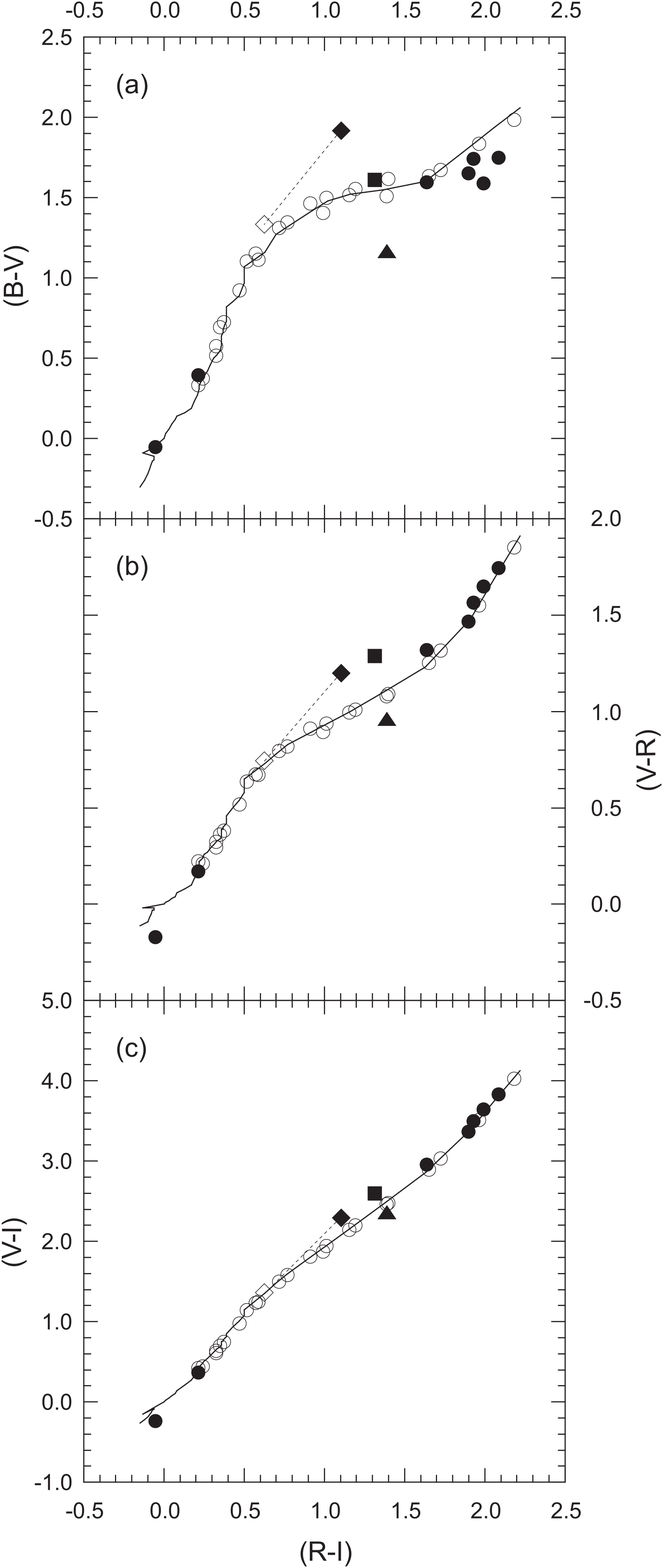}
\caption{Synthetic (a) $(B-V)$/$(R-I)$, (b) $(V-R)$/$(R-I)$ and
(c) $(V-I)$/$(R-I)$ colour-colour diagrams for $\epsilon$ Cha
cluster stars (filled symbols) and main-sequence dwarfs (open
circles). In each panel, the solid lines are the locus of
main-sequence stars from spectral types B0 -- M6, derived from
dwarf photometric sequences. Filled diamonds represent the
observed colours of HD104237E, whereas open diamonds connected by
dotted lines show extinction-corrected colours assuming $A_{V}$ =
1.8 mag. Filled squares represent 2MASS J12014343-7835472 which
might be observed through an edge-on disk. Filled triangles
represent HD104237D, which shows some residual blue excess due to
contamination from its nearby Herbig Ae primary, HD104237A. The
uncertainty in the synthetic colours is comparable to the symbol
size.} \label{Fig2}
\end{center}
\end{figure}

\subsection{Temperature-sensitive narrow-band continuum and spectral line indices}

Each spectral line stores information about the pressure (surface
gravity), effective temperature and chemical abundance
(metallicity) of a star, important parameters in understanding
stellar structure and evolution. However, it is not
straightforward to obtain individual physical properties because
in each line all of these properties are intermixed and affect the
strength of these features together.

In this section, we consider temperature-sensitive narrow-band
continuum and spectral line indices which ought to give
correlations comparable to broadband colours, since these are
usually interpreted astrophysically as a temperature sequence. We
adopted six temperature-sensitive indices and list them in Table
2. Three are 'pseudo-continuum' (PC) spectral ratios that measure
the average continuum levels of the spectrum across restricted
wavelength intervals in a manner comparable to narrow-band
photometry (Mart\'in et al. 1996, 1999), and three are
temperature-sensitive spectral line indices defined from CaH
(Reid, Hawley \& Gizis 1995), VO (Mart\'in et al. 1999) and TiO
(Kenyon et al. 1998) molecular band flux ratios.

Figs 3 and 4 compare these six indices for $\epsilon$ Cha cluster
stars and main-sequence dwarfs from the Gliese catalogue as a
function of their $(R-I)$ broadband colour. Solid-lines in each
panel are a consequence of fitting the dwarf data with a low-order
polynomial in order to define the dwarf sequence. In both figures,
we see that the three PC indices and the three molecular spectral
indices are well-matched between the two groups of stars, except
sometimes for 2MASS J12014343-7835472, HD104237D and HD104237E for
reasons that were discussed in Section 3.1.

\begin{table}
\centering \caption{Adopted narrow-band and spectral line indices,
with all wavelengths in Angstrom units.  For the $[$TiO$]_{2}$
index, the bandpass is 30 \AA\, centred at each wavelength.}
\label{Table2} \vspace{0.3cm}
\begin{tabular}{@{}ll@{}}
\hline
Index  & Definition \\
\hline
\multicolumn{2}{l}{Temperature-sensitive indices}\\
PC2    & $F_{7540-7580}$/$F_{7030-7050}$  \\
PC4    & $F_{9190-9225}$/$F_{7540-7580}$  \\
PC6    & $F_{9090-9130}$/$F_{6500-6540}$  \\
CaH2   & $F_{6814-6846}$/$F_{7042-7046}$  \\
VO1    & $F_{7540-7580}$/$F_{7420-7460}$  \\
VO2    & $F_{7990-8030}$/$F_{7900-7940}$  \\
VO     &  VO1 + VO2                       \\
$[$TiO$]_{2}$ & $-2.5$log[$F_{7100}$/($0.8F_{7025}+0.2F_{7400}$)] \\
\\
\multicolumn{2}{l}{Gravity-sensitive indices}\\
NaI    & $F_{8148-8172}$/$F_{8176-8200}$ \\
FeH2   & $F_{9840-9880}$/$F_{9900-9940}$ \\
\\
\multicolumn{2}{l}{Temperature index for A-type stars}\\
CaII   & $F_{3917-3925}$/$F_{3925-3942}$ \\
H$\delta$ & $F_{4062-4074}$/$F_{4074-4126}$ \\ \hline
\end{tabular}
\end{table}

\begin{figure}
\begin{center}
\vspace{-2cm}
\includegraphics[width=100mm]{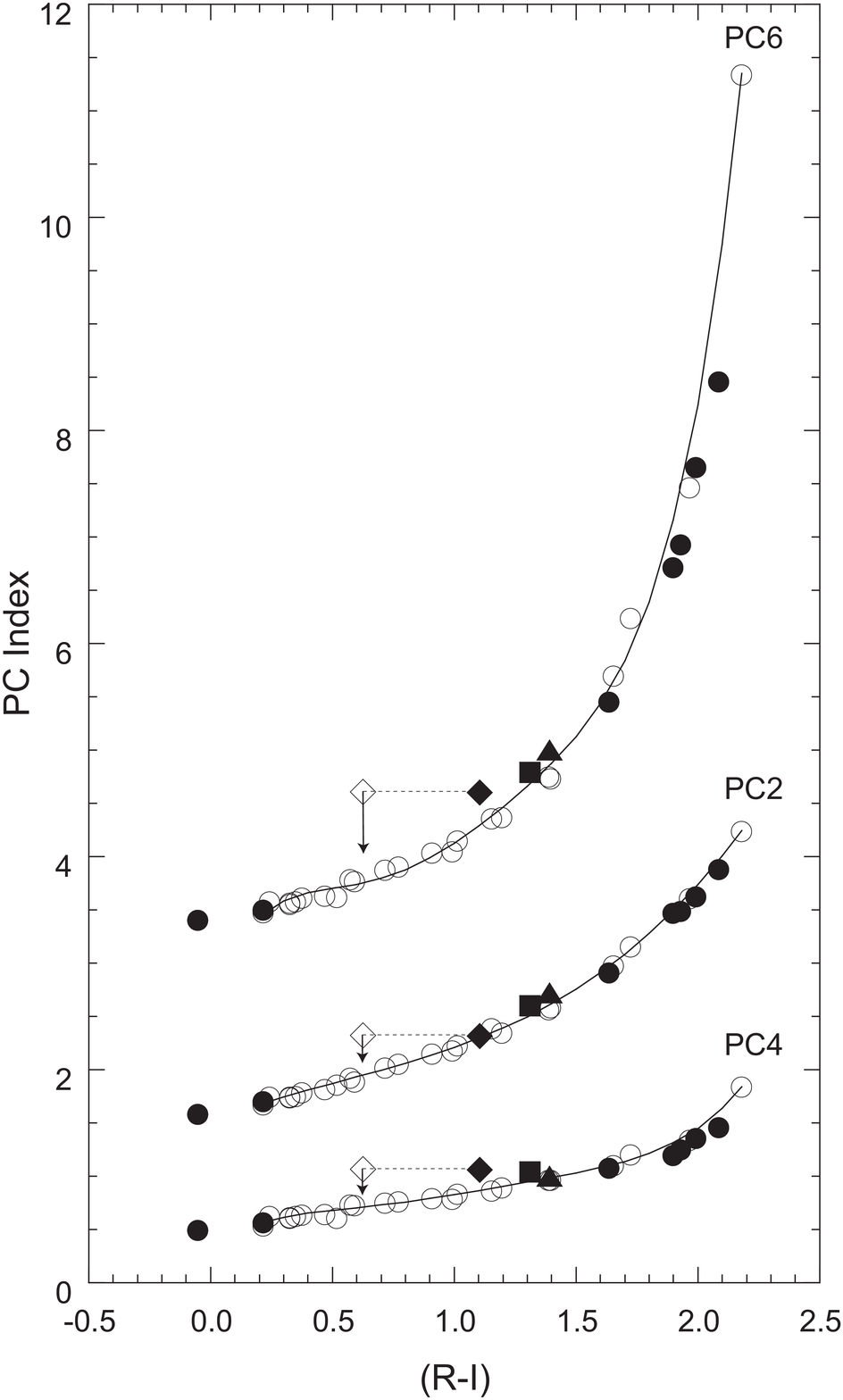}
\vspace{-2.5cm} \caption{PC2, PC4 and PC6 narrow-band spectral
indices for $\epsilon$ Cha stars (filled symbols; see the caption
for Fig. 2 for discussion) and main-sequence dwarfs (open
circles). Open diamonds represent the dereddened $(R-I)$ colour of
HD104237E, with the location of star expected to translate
vertically in the direction indicated by the arrows. The solid
lines are low-order fits to the dwarf data. The uncertainty in the
indices is comparable to the symbol size.} \label{Fig3}
\end{center}
\end{figure}

\begin{figure}
\begin{center}
\includegraphics[width=65mm]{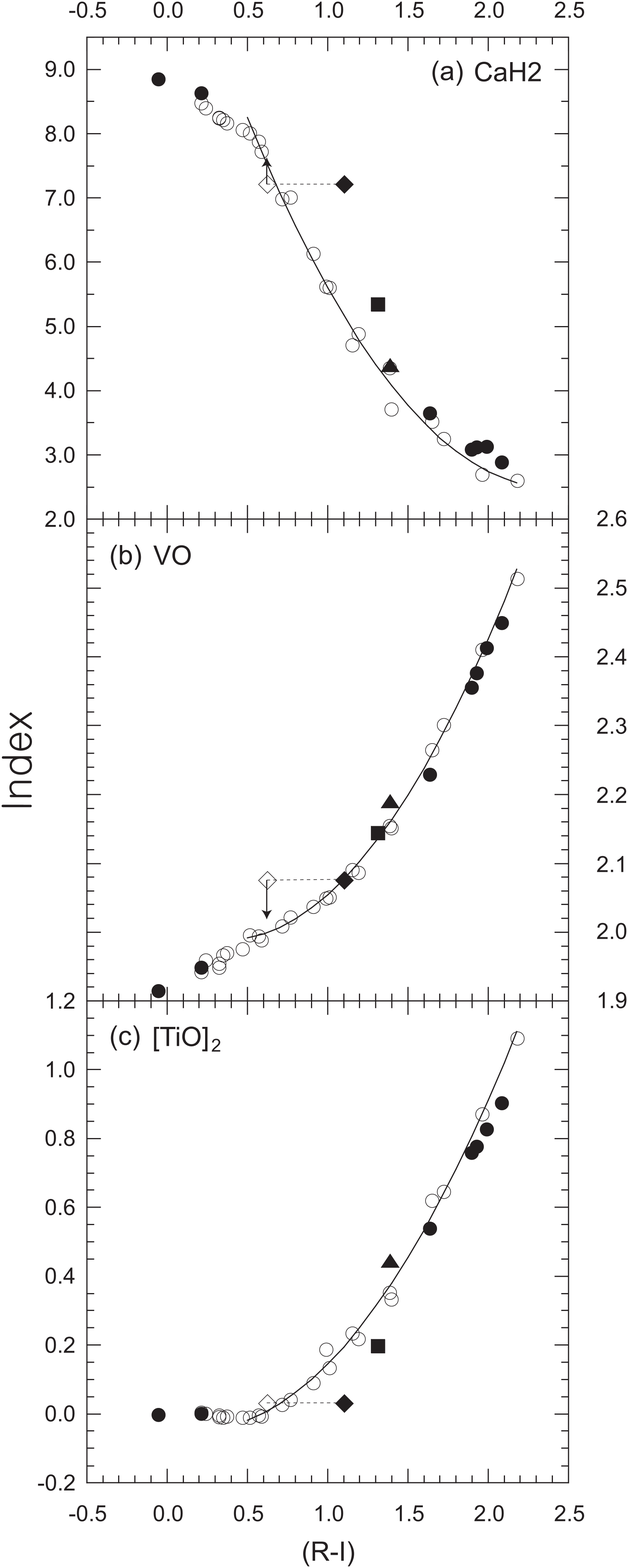}
\caption{(a) CaH2, (b) VO and (c) [TiO]$_{2}$ spectral indices for
the $\epsilon$ Cha cluster (filled symbols; see the caption for
Fig. 2 for discussion), and main-sequence stars (open circles).
Open diamonds represent the dereddened $(R-I)$ colour of
HD104237E, with the location of star then expected to translate
vertically in the directions indicated by the arrows. The solid
lines are low-order fits to the dwarf data for colours greater
than $(R-I)=0.5$. The uncertainty in the indices is comparable to
the symbol size.} \label{Fig4}
\end{center}
\end{figure}

\subsection{Spectral types of the late-type stars}

Table 3 lists the synthetic colours for the late-type stars, along
with spectral types derived from the $(R-I)$ colour (Fig. 2), the
average of the PC ratios (Fig. 3), and the CaH, VO and TiO
molecular band strengths (Fig. 4) in comparison to the dwarf
sequence provided by observations of Gliese disk dwarfs. The
adopted spectral types for these stars are average values.
Overall, there is good agreement between the different temperature
indices, with scatter for most stars at the $\pm 0.2$ sub-type
level. The exceptions are 2MASS J12014343-7835472 (star with a
possible edge-on disk) and HD104237E (star with significant
reddening). For 2MASS J12014343-7835472 we adopt a spectral type
based on molecular line information only, not from the broad- and
narrow-bands colours since their continua are affected by the
presence of the disk. For HD104237E, the adopted K2.5 spectral
type is the average of spectral estimates given by Feigelson et
al. (2003) and Grady et al. (2004). For the other star with some
unusual colours, HD104237D, the temperature-sensitive indices
appear unaffected as we see no evidence that the spectrum is
contaminated at wavelengths beyond 6000\AA\, by the nearby bright
HD104237A.

\begin{table*}
\centering \caption{Synthetic colours for the $\epsilon$ Cha
cluster stars derived from the low-resolution DBS spectra.
Spectral types are derived from comparison with dwarfs in the
broadband {\it BVRI\,} colours, represented here by the $(R-I)$
colour, the average of the three pseudo-continuum (PC) spectral
ratios and three line indices representing the CaH, VO and TiO
molecular bands strengths. For the two early-type systems,
$\epsilon$ ChaAB and HD104237A, we derived their spectral types
using the strength of CaII and H$\delta$ spectral lines. HD104237E
is extinguished by $A_{V} = 1.8$ mag. The colours listed are those
measured without correction for extinction. For 2MASS
J12014343--7835472, the values in brackets are obtained after
removal of the strong H$\alpha$ emission line from the spectrum
and recalculation of the indices. The final column of the table
lists our adopted spectral types for these stars; see Sections 3.3
and 3.4 for details.} \label{Table3} \vspace{0.3cm} \scriptsize{
\begin{tabular}{@{}lcccccccccccrc@{}}
\hline
& \multicolumn{4}{c}{Synthetic Colours} && \multicolumn{7}{c}{Spectral Types} \\
Star&$(B-V)$&$(V-R)$&$(V-I)$&$(R-I)$&&$(R-I)$&PC&CaH2&VO&[TiO]$_{2}$&CaII(H$\delta$)&Adopted\\
\hline
$\epsilon$ ChaAB &--&--&--&--&&--&--&--&--&--&B9&B9\\
HD 104237A &--&--&--&--&&--&--&--&--&--&A4(A5)&A4/5\\
HD 104237E  & 1.92:&1.20:&2.31:&1.10:&&--&--&K5.7:&M1.4:&K6.6:&--&K2.5\\
2MASS J12014343--7835472 &1.62:&1.40(1.29):&2.59(2.61):&1.19(1.31):&&--&--&M1.2:&M2.9:& M1.6:&--&M2:\\
HD 104237D & 1.17:&0.97:&2.37:&1.39&&M3.0&M3.3&M2.9&M3.3&M3.4&--&M3.2\\
2MASS J12074597--7816064 &1.60&1.32&2.97&1.63&&M4.1&M3.9&M3.3&M3.8&M3.8&--&M3.8\\
CXOU J115908.2--781232 &1.66&1.47&3.38&1.89&&M5.0&M4.6&M4.6&M4.8&M4.8&--&M4.7\\
CXOU J120152.8--781840 &1.75&1.57&3.51&1.92&&M5.1&M4.8&M4.4&M4.9&M4.8&--&M4.8\\
USNO-B120144.7--781926 &1.60&1.65&3.66&1.99&&M5.2&M5.2&M4.4&M5.2&M5.0&--&M5.1\\
2MASS J12005517--7820296 &1.76&1.75&3.85&2.08&&M5.5&M5.3&M4.8&M5.4&M5.3&
--&M5.3\\
\hline
\end{tabular}}
\end{table*}

\subsection{Spectral types of the early-type stars}

For the two early-type systems in the cluster, $\epsilon$ Cha AB
and HD104237A, we adopted line indices derived from the
$\lambda$3934\AA\, Ca II and $\lambda$4102\AA\, H$\delta$
absorption lines as temperature-sensitive and therefore spectral
type indicators (Hern\'andez et al. 2004). Table 2 lists the
working definition of each index.

Instead of plotting the indices, we show the spectra of the stars
stacked by derived spectral type. Fig. 5 shows low-resolution
spectra ($\lambda\lambda 3750 - 4150$) obtained from blue arm DBS
observations of the two early-type stars in the $\epsilon$ Cha
cluster ($\epsilon$ ChaAB and HD 104237A), the three early-type in
the $\eta$ Cha cluster ($\eta$ Cha, HD 75595 and RS Cha), five B-
and A-type stars listed in the Bright Star Catalogue, and four
F-type stars from the Gliese Catalogue. We can clearly see that
the strength of Ca II absorption line becomes stronger with later
spectral type (decreasing temperature), while the strength of
H$\delta$ becomes weaker. However, we need to exercise caution
when estimating the spectral types for Herbig stars, since these
lines can be filled with emission. We discuss the effect of
emission in-fill further below.

$\epsilon$ ChaAB is a binary system consisting of a B9 primary and
an early A-type secondary with separation variously reported
between 0.45$''$ and 1.9$''$ (McAlister et al. 1990; ESA 1997;
Worley \& Douglass 1997; Dommanget \& Nys 2002). The primary star
is not a Herbig Be star even though it is likely coeval with the
other early-type cluster member, the Herbig Ae star HD104237 (Hu
et al. 1991; Knee \& Prusti 1996; Eggen 1998). Accordingly we
assume there is no contamination of the Ca II and H$\delta$ lines
due to emission, and we can confirm the published spectral type,
B9, for this star by comparison with those of standards.

There has been a discrepancy in the spectral type of HD 104237A,
perhaps a consequence of different spectral typing methods.
Published spectral types include: A4IVe (Hu et al. 1991), A4 --
A5Ve (Blondel et al. 1993), A7IVe (Brown et al. 1997), and A7.5 --
A8Ve (Grady et al. 2004). In deriving a spectral type for this
object using Ca II and H$\delta$ absorption lines we should
consider the possibility of contamination from emission. Other
spectral features are clearly in emission; Fig. 1 shows emission
in H$\alpha$ and the Ca II triplet (8498 \AA\,, 8542 \AA\,, 8662
\AA\,). However, for A-type stars such as HD104237A the reverse
relationship in the strength of Ca II and H$\delta$ with
increasing spectral type (decreasing temperature) enables us to
constrain the spectral type of this star, even if these lines are
partly filled by emission. Contamination by emission of the Ca II
line would lead to the deduction of a earlier spectral type since
the equivalent width of the line increases with increasing
spectral type, whereas contamination by emission of H$\delta$
would lead to a later spectral type since the equivalent width of
the line decreases with increasing spectral type. For HD104237A we
obtained a spectral type of A4 from the Ca II line and A5 from the
H$\delta$ line, suggesting that emission is not significantly
distorting the profiles of these lines. Therefore we adopt a
spectral type of A4/A5 for this star, and note that this result is
consistent with the majority of the spectral types assigned by
previous studies.

\begin{figure}
\begin{center}
\includegraphics[width=70mm]{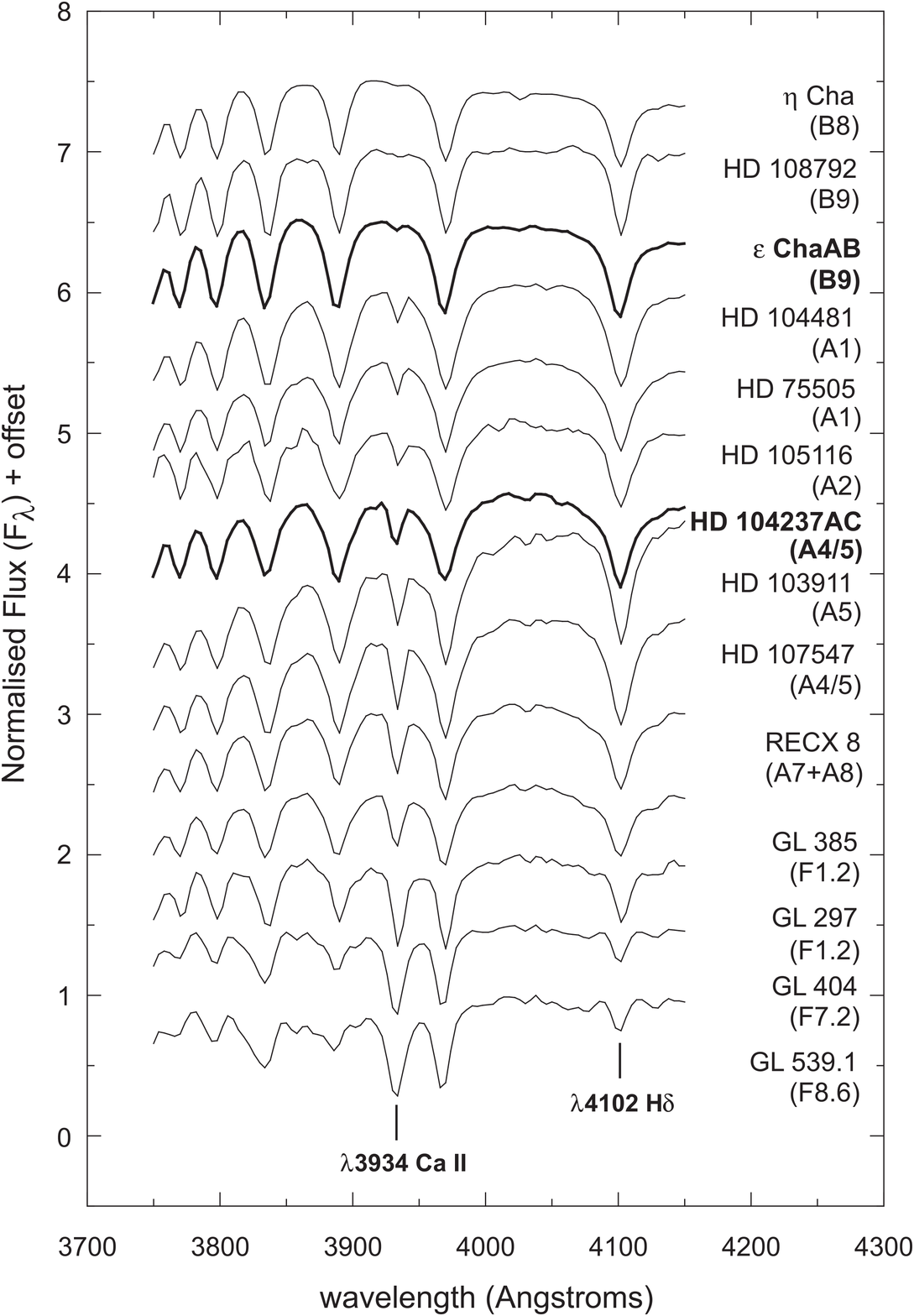}
\caption{Low-resolution blue spectra ($\lambda\lambda 3750-4150$)
of the two early-type systems in the $\epsilon$ Cha cluster
($\epsilon$ ChaAB and HD 104237A; thick lines), the three
early-type systems from the $\eta$ Cha cluster ($\eta$ Cha, HD
75505 and RS Cha AB), along with five B- and A-type stars from the
Bright Star Catalogue, and four F-type stars from the Gliese
Catalogue. We ordered the spectra by the strength of the
$\lambda3934$ Ca II line and the inverse-strength of the
$\lambda4102$ H$\delta$ line. These are sensitive indicators of
stellar spectral type for late-B to F-type stars.} \label{Fig5}
\end{center}
\end{figure}

\section{Gravity trend with age}

From comparison of the broadband synthetic colours ($VRI$),
narrowband continuum (PC indices), and molecular spectral bands
(CaH, VO and TiO) between the young $\epsilon$ Cha cluster and old
disk dwarfs, we found the spectroscopic properties of these two
groups of stars to be essentially identical; see Figs 2, 3 and 4
and their associated text. However, in detail we see the
difference in gravity sensitive atomic- and molecular lines
between these two groups of objects. Fig. 6 shows the difference
in the strength of the $\lambda$8183, 8195 Na I doublet and
$\lambda$9896 FeH molecular band for the $\epsilon$ Cha cluster,
the $\eta$ Cha cluster, disk dwarfs, and M giants. Table 2 lists
the working definition for each index.

For both gravity-sensitive indices, the distribution for
$\epsilon$ Cha stars, like that for $\eta$ Cha stars, occupies the
region between well-defined dwarf and giant sequences. This
indicates, for both these PMS groups, intermediate surface
gravities between those of giants and dwarfs. Amongst the late-M
stars, there is also a clear tendency for $\epsilon$ Cha stars to
have weaker gravity indicators than $\eta$ Cha stars of similar
spectral type. This result is consistent with the $\epsilon$ Cha
stars being a younger PMS group than $\eta$ Cha.

We provide a visual comparison of the gravity-sensitive features
in Fig. 7, where we compare spectra of two late-type $\epsilon$
Cha and two $\eta$ Cha cluster members of similar spectral type,
as determined from their broad-band colours and
temperature-sensitive indices. The comparison clearly shows that
the strength of gravity-sensitive spectral lines such as the
$\lambda$8183, 8195 Na I doublet and $\lambda$9896 FeH is weaker
in the younger $\epsilon$ Cha stars compared to the $\eta$ Cha
stars.

\begin{figure}
\begin{center}
\includegraphics[width=75mm]{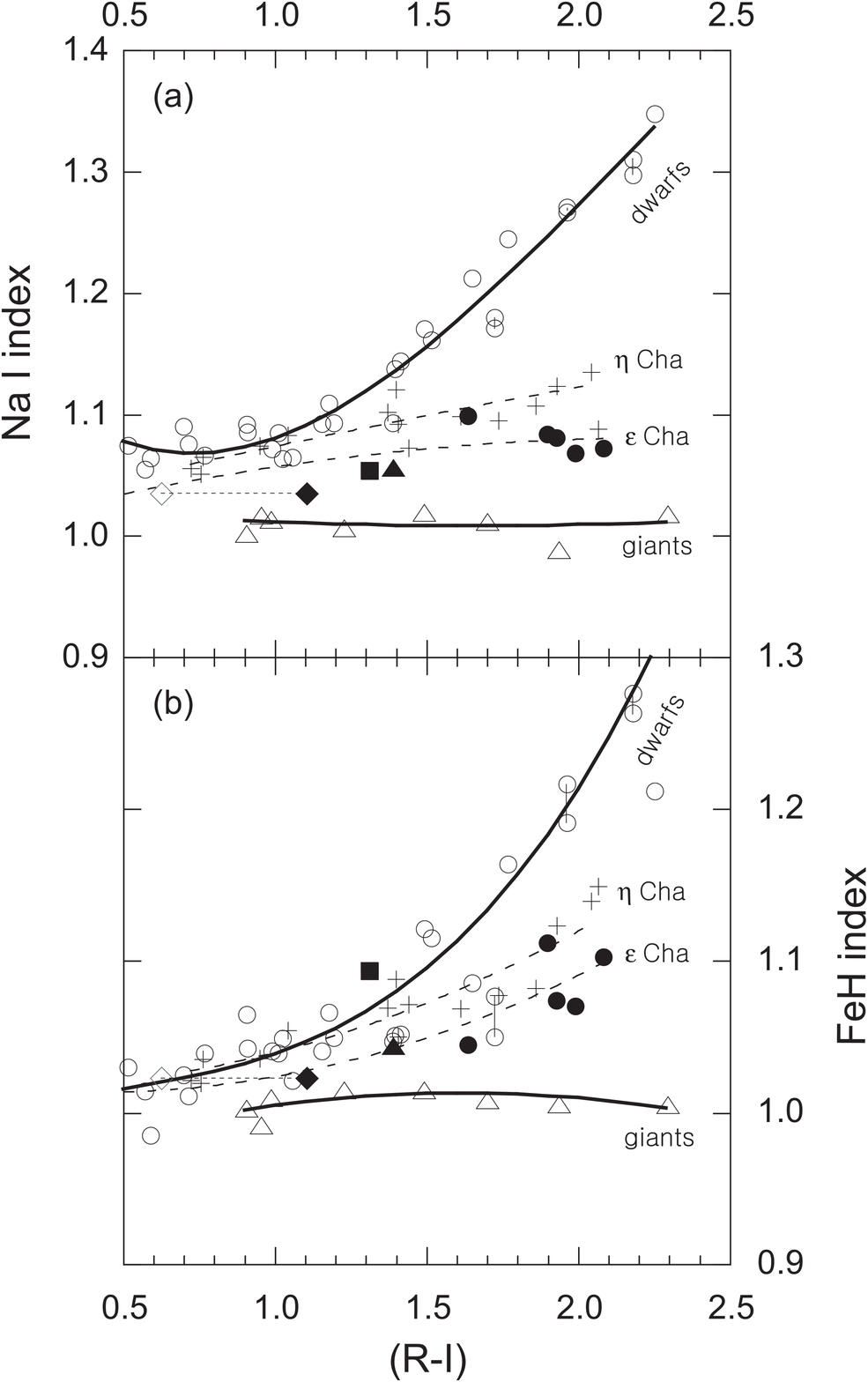}
\caption{Comparison of gravity-sensitive indices that measure the
strength of (a) the $\lambda 8183, 8195$ Na I doublet, and (b) the
$\lambda 9896$ FeH band for $\epsilon$ Cha cluster stars (filled
symbols, with open diamonds representing the deredenned
HD104237E), compared to stars in the $\eta$ Cha cluster (crosses),
old disk dwarfs (open circles) and giants (open triangles). The
three connected pairs of points in the panels represent the disk
dwarfs GL 514.1, GL 473 and GL 406 which were observed in DBS runs
in 2002 (Lyo et al. 2004) and in 2005 (this paper). The small
differences of the indices derived from different DBS spectra
provides an indication of the uncertainty of individual
observations. The dashed lines in both panels are low-order fits
to the $\eta$ Cha and $\epsilon$ Cha observations.} \label{Fig6}
\end{center}
\end{figure}

\begin{figure}
\begin{center}
\includegraphics[width=80mm]{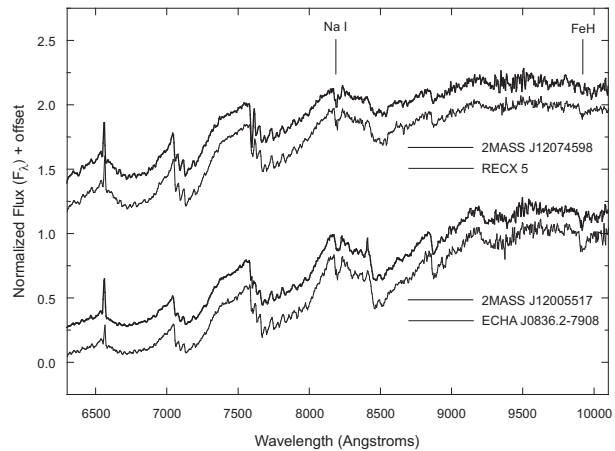}
\caption{Comparison of late-type $\epsilon$ Cha cluster and $\eta$
Cha cluster members (Lyo et al. 2004) of similar spectral type:
(upper) 2MASS J12074597--7816064 and RECX 5 (spectral type M3.8)
and (lower) 2MASS J12005517--7820296 and ECHA J0836.2--7908
(spectral type M5.3). Gravity-sensitive spectral lines such as Na
I and FeH are weaker in the younger $\epsilon$ Cha stars. }
\label{Fig7}
\end{center}
\end{figure}

\section{Blue excess in low-mass PMS stars}

Late-M stars in the $\epsilon$ Cha cluster show a $(B-V)$ colour
excess of $\approx 0.2$ mag compared to observations of Gliese
stars and the dwarf colour sequence; see Fig. 2(a). This result is
consistent with that seen in late-M stars in the $\eta$ Cha
cluster (Lyo et al. 2004; see their fig. 2), suggesting that it
might be a general property of young stellar objects.

Fig. 8 shows that the flux excess is in the $U$-band (centred at
$\lambda 3600$ \AA) and $B$-band (centred at $\lambda 4500$ \AA)
and not the result of a flux deficit in the $V$-band (centred at
$\lambda 5500$ \AA). In Fig. 8(a) we compare spectra of the
$\epsilon$ Cha cluster member USNO-B120144.7--781926 $(R-I =
1.99)$ and a Gliese main-sequence dwarf of similar spectra type,
GL 473 $(R-I = 1.96)$. In Fig. 8(b), where we show the ratio of
these spectra after normalizing them at $\lambda 5300$ \AA, we see
that the main source of the blue excess is enhanced blue
continuum, though the spectrum of the cluster member shows
prominent Balmer and Ca II H and K emission as well. Despite the
presence of these strong blue emission lines, the star shows a
typical level of $B$-band excess in Fig. 2(a).

A blue excess has been documented in Pleiades K-type stars for
over 40 years (Herbig 1962; Jones 1972; Landolt 1979; Stauffer
1980; van Leeuwen et al. 1987) and in G- to K-type stars in other
intermediate-aged open clusters, such as $\alpha$ Persei, NGC
2516, IC 2602 and $\delta$ Lyrae. Combined results for the $\sim
5$ Myr-old $\epsilon$ Cha cluster (this paper), $\approx 8$
Myr-old $\eta$ Cha cluster (Lyo et al. 2004), $\sim 125$ Myr-old
Pleiades (Stauffer et al. 2003) and the $\sim 160$ Myr-old NGC
2516 (Sung et al. 2002), show a tendency for increasing blue
excess with decreasing stellar mass/temperature, i.e. a 0.1 mag
excess for K- to early M-type stars increasing to a 0.2 mag excess
for late-M stars. However, for stars of a given type there appears
to be no significant dependence on the level of excess as a
function of stellar age in systems that have been studied with
ages between several Myr and $\sim 200$ Myr.

The cause of the blue excess in young stellar populations has been
suggested to be the result of enhanced chromospheric activity
(Sung et al. 2002), or alternatively the spottedness of stars
resulting from convection and rapid rotation (Stauffer et al.
2003). We consider each of these possibilities in turn.

Enhanced X-ray activity exceeding the solar level by several
orders of magnitude (measured as a function of the fractional
X-ray luminosity $L_{X}/L_{\rm bol}$) is a general property of
low-mass PMS stars (Feigelson \& Montmerle 1999; Feigelson et al.
2002; Favata \& Micela 2003; G\"udel 2004). From observations of
stars in the Orion Nebular Cloud, NGC 2264, Chamaeleon I,
Pleiades, Hyades, and Gyr-old disk stars, i.e. stellar populations
ranging in age from 1 Myr to several Gyr, Preibisch \& Feigelson
(2005) found that for M-type stars the $L_{X}/L_{\rm bol}$ ratio
and the X-ray surface flux $F_{X}$ decreases only mildly for the
first $\sim 100$ Myr. The decay in X-ray activity was found to be
more-pronounced for G-, F-, and K-type stars. However, the
maintenance of high X-ray activity levels until $\sim 100$ Myr
suggests that the enhanced chromospheric activity might be the
dominant mechanism responsible for the blue excess, since this
timescale closely corresponds to our finding that the blue excess
is present across the PMS phase from ages of several Myr to a few
hundred Myr.

In much older populations, X-ray activity is seen to have decayed
significantly in the $\sim 700$ Myr-old Hyades cluster and in
Gyr-old Galactic field stars, particularly amongst the higher-mass
objects. This result seems consistent with the non-detection of
the blue excess for K-types stars in the $\sim 700$ Myr-old
Praesepe group (Stauffer et al. 2003). Studying K- and M-type
stars in clusters with ages between 200 Myr and $\sim 1$ Gyr would
be useful to investigate the decay in the blue excess as a
function of stellar mass and X-ray activity.

Light-curve studies of weak-lined T Tauri (WTT) stars of the
Taurus-Auriga cloud region (Bouvier et al. 1993) and the TW Hydrae
Association (Lawson \& Crause 2005) show a increase in photometric
amplitude with decreasing wavelength. However, there is no
observational evidence that starspots induce bluer $(B-V)$ colours
in low-mass PMS stars. The observations of Bouvier et al. suggest
that the $(B-V)$ colour becomes redder due to the presence of cool
starspots, if they are the primary source of the photometric
variability in PMS stars as is the case in non-accreting or
low-level accreting WTT stars. High levels of mass accretion
resulting in variability driven by accretion hotspots as seen in
classical T Tauri stars is a phenomenon of only the youngest PMS
populations. Accretion largely disappears in PMS stars by 10 Myr
and this cannot be a universal solution for the blue excess.

\begin{figure}
\begin{center}
\includegraphics[width=75mm]{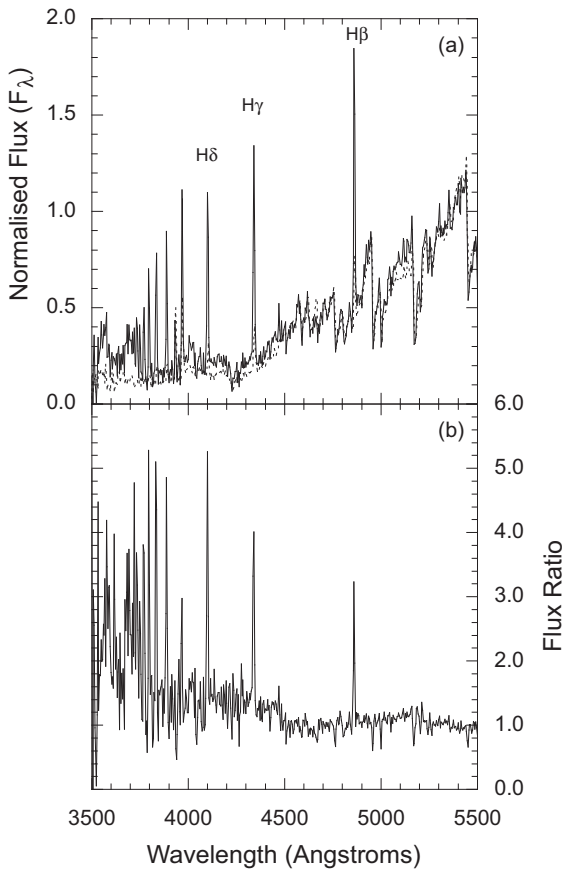}
\caption{(a) Spectra of USNO-B120144.7--781926 [$(R-I)$ = 1.99]
(solid-line) and the standard main-sequence dwarf GL 473 [$(R-I)$
= 1.96] (dotted-line), normalized at 5300\AA  (b) The ratio of the
spectra, obtained by dividing the USNO-B120144.7--781926 spectrum
by the GL 473 spectrum, demonstrates the excess blue emission
present in the PMS star.} \label{Fig8}
\end{center}
\end{figure}

\section{Summary and Conclusions}

Flux-calibrated low-resolution optical spectroscopy is an
indispensable tool for investigating the physical properties of
stars by comparison with other stellar groups and standard star
calibrators. In this paper, we characterized the stellar
population of the $\sim 5$ Myr-old $\epsilon$ Cha cluster via a
number of temperature- and gravity-sensitive spectroscopic
indicators, and compared its properties to the slightly older PMS
group associated with $\eta$ Cha, and to those of Gyr-old disk
dwarfs.

Using synthetic broadband colours, narrow-band continuum, atomic
and molecular line indices derived from the spectra, we find that
the relationship between the broadband colours and spectroscopic
temperature indicators for the $\epsilon$ Cha cluster stars is
indistinguishable from that of the Gyr-old dwarfs. We had
previously reached the same conclusion for the slightly-older
$\eta$ Cha cluster (Lyo et al. 2004). However, there is a clear
evidence that $\epsilon$ Cha cluster stars have lower surface
gravity than $\eta$ Cha cluster stars from measurement of the
gravity-sensitive $\lambda\lambda$8183, 8195 Na I doublet and FeH
molecular spectral lines. This result is consistent with the
$\epsilon$ Cha cluster being slightly younger than $\eta$ Cha, a
few Myr younger according to the HR diagram placement of these two
clusters and comparison with PMS evolutionary model grids.

We also found a $B$-band excess of $\sim 0.2$ mag in the late
M-type cluster members, similar to that found in $\eta$ Cha and in
other PMS populations with ages less than $\sim 200$ Myr. This
result suggests that the blue excess is an ubiquitous property of
low-mass young stellar objects, with the most-likely origin being
enhanced magnetic activity. The presence of significant excess
blue emission appears to closely parallel the phase of high
relative X-ray luminosity seen in low-mass $1-100$ Myr-old PMS
populations.

\section*{Acknowledgments}

ARL and WAL thank the REsearch School of Astronomy and
Astrophysics for the award of telescope time at SSO for this
project. WAL's research is supported by UNSW@ADFA Faculty Research
and Special Research Grants. We thank the referee, Dr R. D.
Jeffries, for making a number of very useful comments on the
paper.

\bsp

\label{lastpage}

\end{document}